\documentclass[pra,reprint,twocolumn,amsmath,amssymb,floatfix]{revtex4-2}
\usepackage{graphicx,comment,color}

\usepackage{subcaption}
\usepackage{physics}
\usepackage{amsmath}
\usepackage{natbib}
\usepackage{xcolor}
\usepackage{ulem}

\begin{document}

\title{Modifying cooperative decay via disorder in atom arrays}
\author{Nik O. Gjonbalaj, Stefan Ostermann, and Susanne F. Yelin}
\affiliation{Department of Physics, Harvard University, Cambridge, Massachusetts 02138, USA}

\date{\today}

\begin{abstract}
	Atomic arrays can exhibit collective light emission when the transition wavelength exceeds their lattice spacing. Subradiant states take advantage of this phenomenon to drastically reduce their overall decay rate, allowing for long-lived states in dissipative open systems. We build on previous work to investigate whether or not disorder can further decrease the decay rate of a singly-excited atomic array. More specifically, we consider spatial disorder of varying strengths in a 1D half waveguide and in 1D, 2D, and 3D atomic arrays in free space and analyze the effect on the most subradiant modes. While we confirm that the dilute half waveguide exhibits an analog of Anderson localization, the dense half waveguide and free space systems can be understood through the creation of close-packed, few-body subradiant states similar to those found in the Dicke limit. In general, we find that disorder provides little advantage in generating darker subradiant states in free space on average and will often accelerate decay. However, one could potentially change interatomic spacing within the array to engineer specific subradiant states.
\end{abstract}

\maketitle

\section{Introduction}

Controllable light-matter interfaces play a central role in the realization of future quantum technologies. They are essential to manipulating and controlling quantum information \cite{Hammerer_2010,Lukin_2003}, for quantum networking applications, and can serve as efficient quantum memories. One promising platform for achieving such interfaces involves arrays of atoms in well-defined geometries that interact with photonic modes of a cavity or waveguide or the electromagnetic vacuum in free space \cite{Barredo_2016, Endres_2016, Barredo_2018}.

While cavities and waveguides can transmit photons over long distances, strong interaction between emitters via coherent photon exchange in free space only occurs if the emitter distance is much smaller than the transition wavelength. The resultant light-induced dipole-dipole interactions \cite{Lehmberg_1970_1,Lehmberg_1970_2} between atoms in such systems are inherently long range, which leads to intriguing cooperative effects. Prominent examples include super- \cite{Dicke_1954,Gross_1982} and subradiance \cite{Guerin_2016,Ferioli_2021_1,Bienaime_2012,Henriet_2019}. Chief among these collective effects are a short-lived superradiant burst of decay and long-lived subradiant modes with suppressed decay rates. Understanding these effects in detail can provide avenues towards efficient photon emission in the superradiant case and robust quantum memory in the subradiant case \cite{Facchinetti_2016,Asenjo-Garcia_2017,Guimond_2019,Rubies-Bigorda_2022,Ballantine_2021,Ferioli_2021_1,Zanner_2022}.

\begin{figure}
    \centering

    \includegraphics[width=8.6cm]{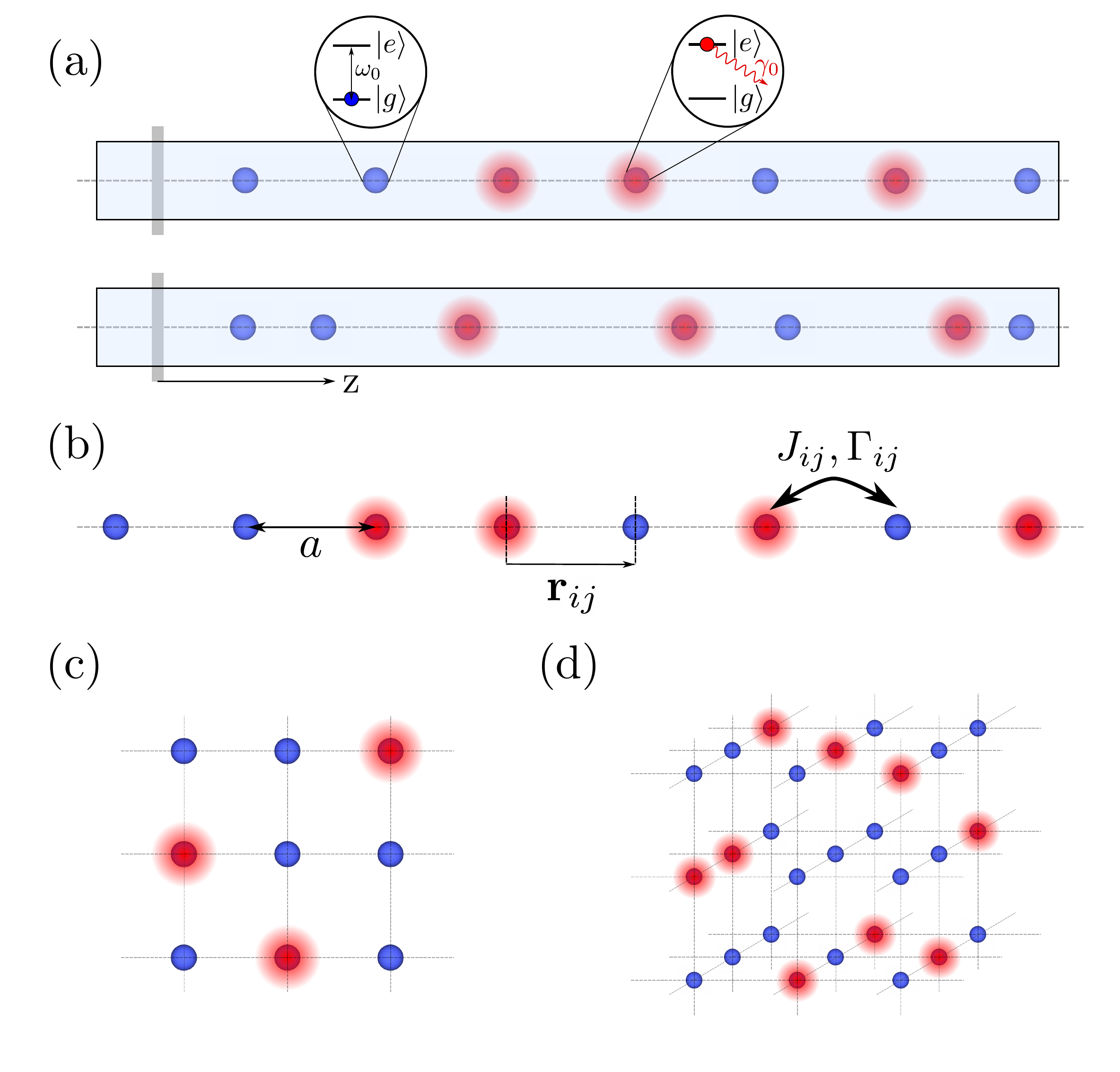}

    \caption{(a) Schematic of a half waveguide with and without spatial disorder. (b-d) Ordered arrays in 1D, 2D, and 3D.}
    \label{fig:schematics}
\end{figure}

The seemingly separate phenomenon of localization within condensed matter physics also provides a method for arresting the relaxation of a many-body initial state. Typically studied in closed systems, the phenomena of Anderson localization \cite{Anderson_1958,Lagendijk_2009} and more recently many-body localization (MBL) \cite{Basko_2006,Gornyi_2005,Alet_2018,Abanin_2019,Nandkishore_2015} describe the halting of transport within a disordered medium of single and multiple interacting excitations, respectively. In such a regime, the system will fail to thermalize and will instead show a strong memory of initial conditions and slow growth of entanglement. Although one might expect that, upon coupling the system to a bath to create open dynamics, any localization would be lost, this is not always the case. This insight has gained much interest in recent years, as reflected in a series of studies on localization or the lack thereof in open systems and more generally the role of disorder in collective decay \cite{Fayard_2021,Yusipov_2017,Fischer_2016,Levi_2016,Nandkishore_2017,Luitz_2017,Luschen_2017,Weidemann_2021,Skipetrov_2014,Skipetrov_2018,Cottier_2018}, showing that different types of localization can be either destroyed or preserved when dissipation is added. In particular, Ref.~\cite{Weidemann_2021} notes that while spectral localization in the spatial support of eigenstates guarantees dynamical localization of energy in closed systems, the same does not apply to open systems. Such distinctions provide a rich phenomenology to study when considering the effect of disorder in open systems where the environment cannot be ignored.

The question of whether Anderson localization of light exists in disordered 3D collections of emitters was previously investigated in Ref.~\cite{Skipetrov_2014}. It was found that fully disordered clouds of atoms did not display localization: although many of the slowest-decaying eigenstates displayed spatial localization over just a few sites, no suppression of decay was found. This led to the conclusion that any slowdown in decay arose because of subradiance and not an analog of Anderson localization. Recently, however, Ref.~\cite{Fayard_2021} showed that many aspects of Anderson localization and MBL persist in chains of atoms confined to a half waveguide, a 1D waveguide with a mirror at one end. In the dilute regime, where the interatomic spacing is greater than the emission wavelength, the system exhibits both spectral localization of eigenstates and dynamical localization of energy, the latter of which halts transport both within the bulk and also from the bulk into the vacuum. In order to bridge these results and expand upon their analyses, we investigate whether disorder can cause spectral and dynamical localization of single excitations in both the half waveguide and 1D, 2D, and 3D arrays of atoms in free space. In contrast to Ref.~\cite{Skipetrov_2014}, we systematically increase disorder, starting from an ordered case, where atoms are assumed to be trapped in a periodic configuration, and consider multiple different disorder strengths until we reach full disorder. We also analyze the role of dimensionality by considering 1D, 2D and 3D configurations. Beyond that, we investigate the role of lattice spacing and density, which goes beyond the analysis in Ref.~\cite{Fayard_2021}. In all cases, we observe that spatial disorder causes spectral localization, reducing the support of the delocalized eigenstates. However, this does not automatically guarantee dynamical localization of energy within the array, i.e. a suppression of decay. While select regions of parameter space show an advantage to using disorder to create more subradiant states, it mostly accelerates decay to the vacuum. We explain these phenomena using dark states occurring in Dicke superradiance~\cite{Dicke_1954,Gross_1982} and back up our interpretations using various numerical results.

The remainder of the paper is outlined as follows. In Sec.~\ref{sec:hwg}, we revisit the half waveguide from Ref.~\cite{Fayard_2021} and expand the analysis of the single-excitation spectrum to both the dilute and dense regimes, laying the groundwork for the phenomena behind the localization. In Sec.~\ref{sec:1d free}, we move to 1D chains of atoms in free space and repeat the analysis. We explain our results through approximate few-body Dicke states and comment on why disorder accelerates or suppresses decay in different regimes. Finally, in Sec.~\ref{sec:higher d arrays}, we move to 2D and 3D arrays and see similar results while commenting on the diminishing returns of disorder.

\section{Half Waveguide}
\label{sec:hwg}

\begin{figure*}
	\centering

	\begin{subfigure}{0.49\linewidth}
		\caption{}
		\includegraphics[width=8.6cm]{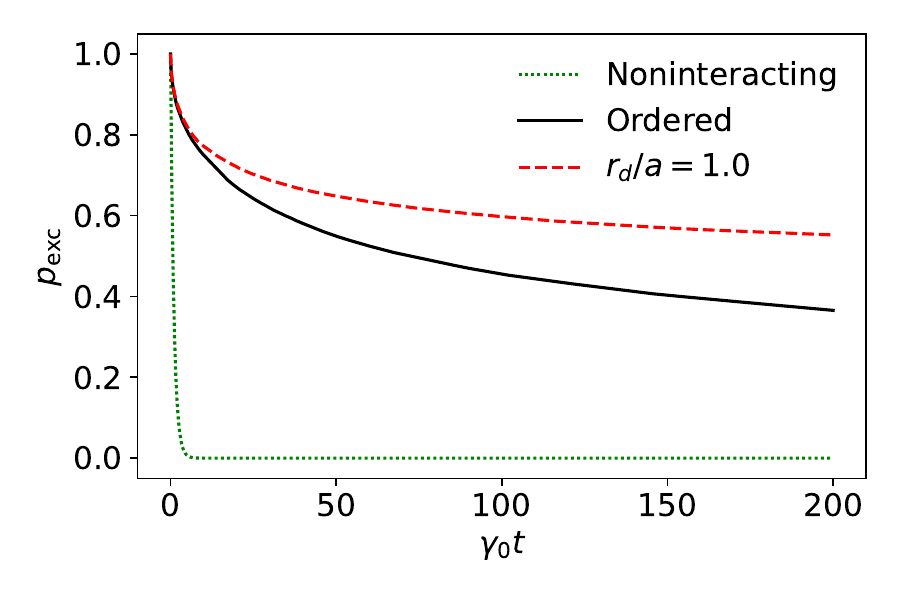}
		\label{fig:hwg time evol}
	\end{subfigure} \hfill
    \begin{subfigure}{0.49\linewidth}
		\caption{}
		\includegraphics[width=8.6cm]{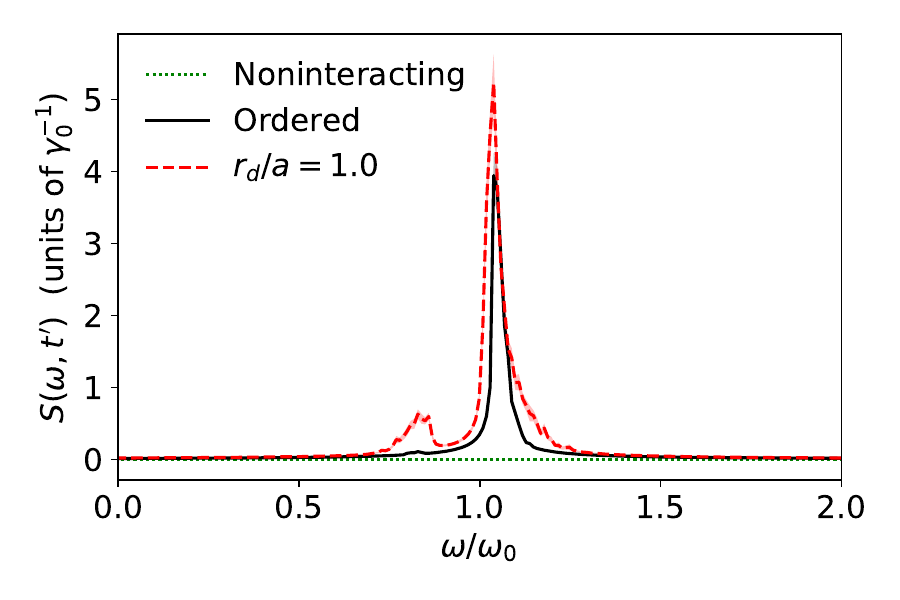}
		\label{fig:hwg flourescence spectrum}
	\end{subfigure}

	\caption{(a) Time evolution of excited atom population in a randomly-initialized $N=50$ half waveguide. The ordered case (black) already displays subradiance relative to the noninteracting case (green), while static disorder (red) further slows the decay. (b) Dynamic fluorescence spectrum of evolution in (a) evaluated at $\gamma_0 t' = 100$.}
	\label{fig:hwg intro}
\end{figure*}

\begin{figure}
	\centering
	
	\begin{subfigure}{\linewidth}
		\caption{}
		\includegraphics[width=8.6cm]{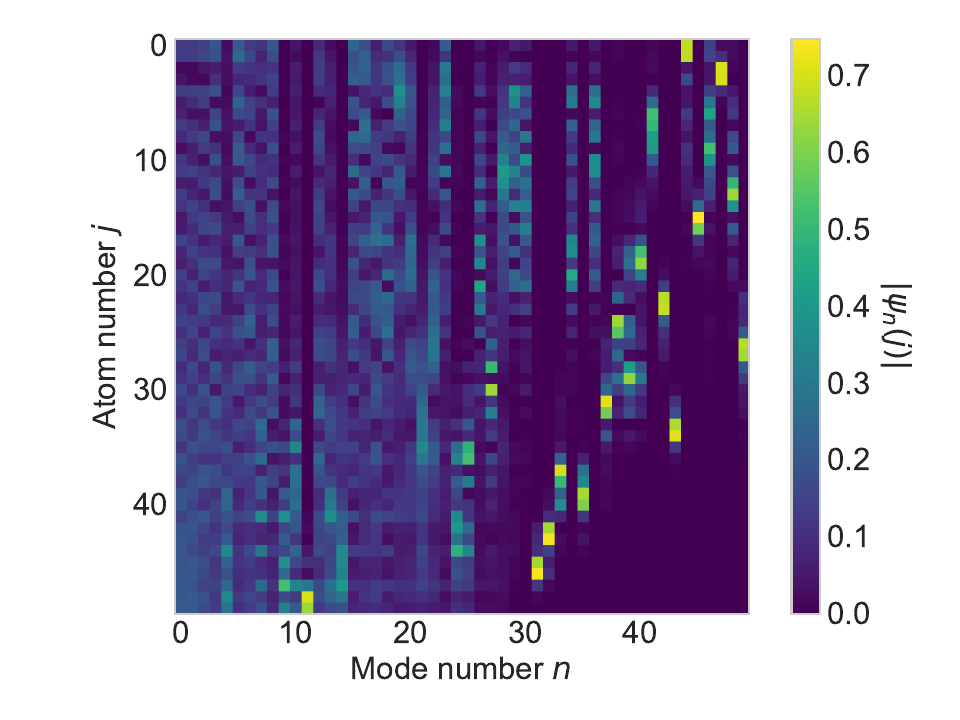}
		\label{fig:hwg eigenmodes}
	\end{subfigure}
	
	\begin{subfigure}{\linewidth}
		\caption{}
		\includegraphics[width=8.6cm]{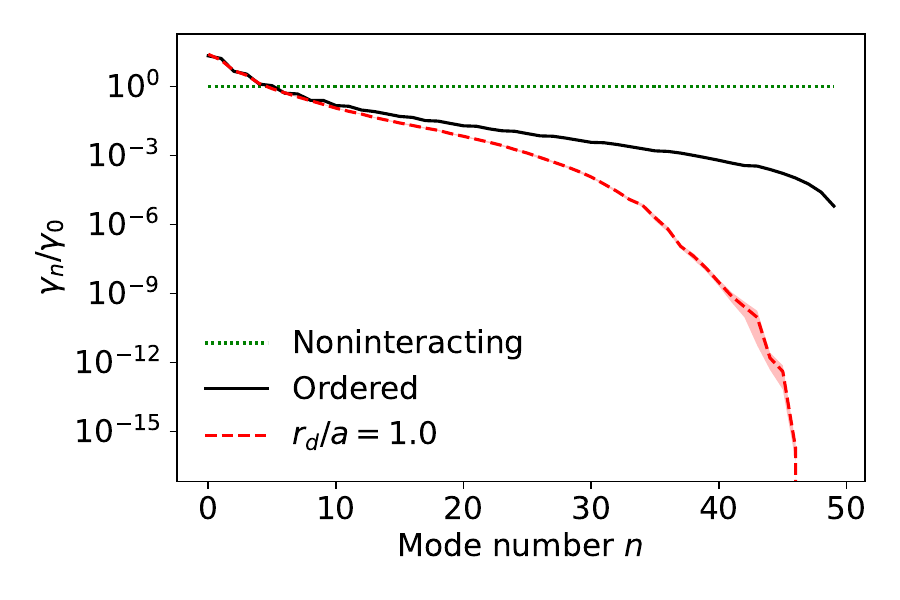}
		\label{fig:hwg decay rates}
	\end{subfigure}

	\caption{The spatial profile $|\psi_n(j)|$ (a) and decay rates $\gamma_n = -2 \: \mathrm{Im}[E_n]$ (b) of single-excitation eigenmodes of the $N=50$ half waveguide non-Hermitian Hamiltonian, ordered by their decay rate  from fastest (left) to slowest (right) decay. In (a), we plot the eigenmodes of a single realization of spatial disorder, while in (b) we compare the rates of a single noninteracting atom, the full ordered chain, and the disordered chain averaged over realizations. We consider spatial disorder with maximum strength $r_d = a$. The thin ribbon corresponds to the standard deviation in the mean.}
	\label{fig:hwg modes}
\end{figure}

\begin{figure}
	\centering
	\includegraphics[width=8.6cm]{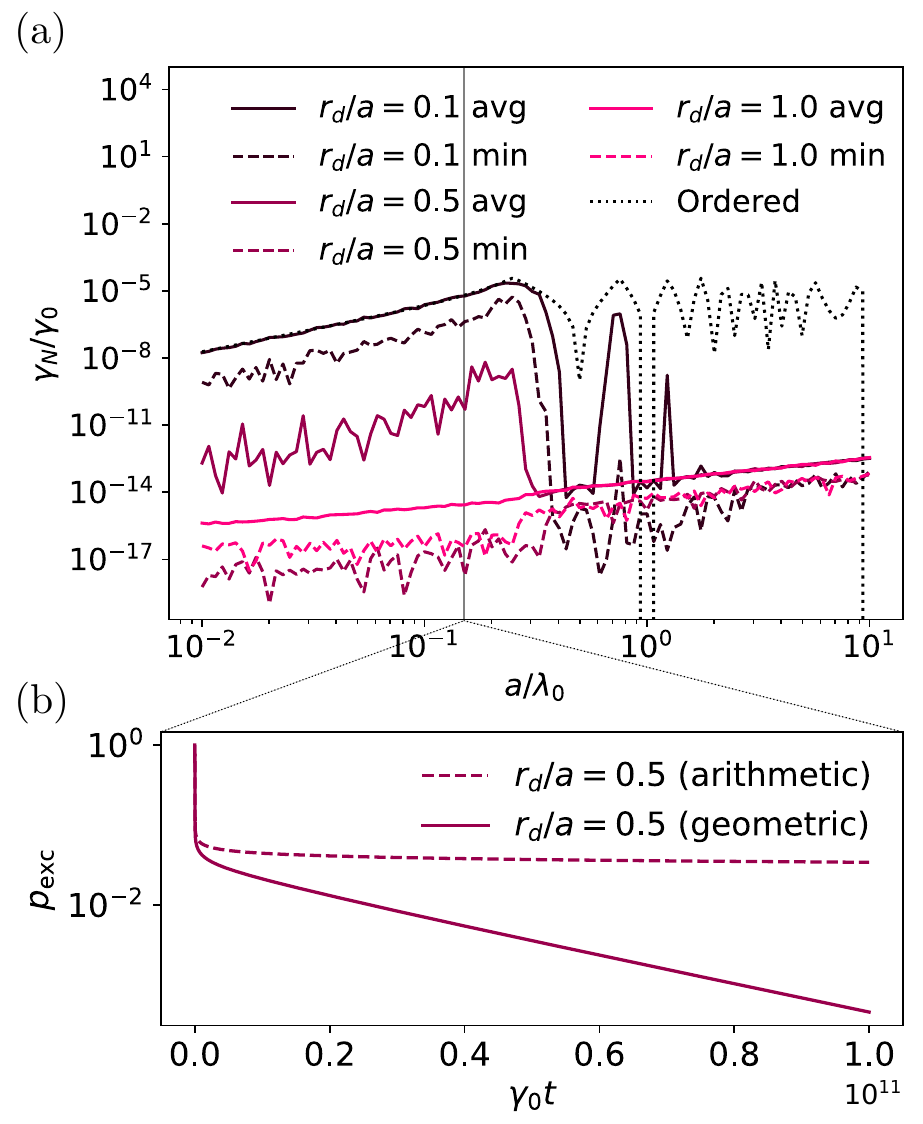}
	\caption{(a) Average (thick) and minimum (thin) magnitude of the slowest decay rate $\gamma_N$ in the spectrum of an $N=50$ half waveguide as a function of lattice spacing for 3 disorder strengths. At late times, the average decay rate determines the geometric mean of the excited population, whereas the minimum dominates the arithmetic mean. An example for $r_d/a = 0.5$ and $a/\lambda_0 = 0.15$ is shown in (b).}
	\label{fig:hwg lattice spacing}
\end{figure}

The analysis in this section is strongly inspired by Ref.~\cite{Fayard_2021}. We not only expand upon their results but also use it as a guideline for the analysis of the free space case presented below. In a half waveguide, pictured in Fig.~\ref{fig:schematics}a, an ensemble of $N$ two-level atoms with transition frequency $\omega_0$ interacts with a one-dimensional continuum of modes. Because the speed of light $c$ is by far the shortest time scale in the system, we can use the Born-Markov approximation and trace out the photonic degrees of freedom. This results in a dissipative ``spin'' model described by the following master equation ($\hbar = 1$ for the remainder of this work) \cite{Fayard_2021,Chang_2012,Le_Kien_2005,Lalumiere_2013}

\begin{align}
    \dot{\rho} = - i \left[ H_{\mathrm{hwg}} \rho - \rho H_{\mathrm{hwg}}^{\dagger} \right] + \sum_{ij} \Gamma_{ij} \sigma_i^{ge} \rho \sigma_j^{eg} \: , 
\end{align}
where
\begin{align}
    H_{\mathrm{hwg}} &= \omega_0 \sum_{i} \sigma_i^{ee} - i \frac{\gamma_0}{2} \sum_{i j} \Big[ \exp\left( - i k_0 |z_i - z_j|  \right) \\ \nonumber
    &- \exp\left( - i k_0 |z_i + z_j|  \right) \Big] \sigma_i^{eg} \sigma_j^{ge} \: .
    \label{eq:H hwg}
\end{align}
Here, $\Gamma_{ij} = \gamma_0 \left[ \cos(k_0|z_i - z_j|) - \cos(k_0|z_i + z_j|) \right]$, $k_0 = \omega_0/c$, and $\gamma_0$ is the spontaneous emission rate of a single excited atom in a full waveguide with no mirror. In addition, $\sigma_i^{eg} = \ket{e_i}\bra{g_i}$ and $\sigma_i^{ge} = \ket{g_i}\bra{e_i}$ are the atomic transition operators, which raise and lower the spin state of the $i$th atom. Throughout this work, we focus on the single-excitation regime, which implies that the on-site energies only contribute an overall phase to the wavefunction and therefore only appear as an overall shift when calculating the dynamic fluorescence spectrum below.


Given this model, we numerically analyze the effect of disorder on the system. Although analytics can be used to exactly solve few-atom systems, we find that our results can in general be understood using the solutions to the simplest such case: a two-atom system. This setup has been solved and analyzed throughout the literature (e.g. \cite{Lehmberg_1970_2}). Increasing the number of atoms $N$ in analytic calculations adds complexity without providing a much deeper understanding of our current results; as such, we focus on numerics in this work. Simulating the entire master equation significantly limits our system sizes to about $N \lesssim 10$, so we consider some alternatives. Reference~\cite{Rubies_Bigorda_2023} uses a cumulant expansion of the Heisenberg-Langevin equations up to 3-body cumulants. Although this method gives accurate simulations for large systems at short times, it begins to give unphysical results as time goes on and 4-body cumulants become important. Because we are analyzing long-lived subradiant states and searching for even slower decay, we found the cumulant expansion inapplicable in this regime. Rather, we restrict ourselves to the $N_{\mathrm{exc}} = 1$ manifold to look for analogs of Anderson localization. In this limit, the term $\propto\sigma_i^{ge} \rho \sigma_j^{eg}$ won't contribute to dynamics in the single-excitation subspace \cite{Needham_2019}, and the entire time evolution is governed by the non-Hermitian Hamiltonian $H_{\mathrm{hwg}}$.


We begin by analyzing dynamics in an $N=50$ chain of atoms coupled to a 1D half waveguide. To avoid any bias in the choice of initial conditions, we use an equal superposition of all singly-excited basis states and multiply each such state by a random phase. Ref.~\cite{Fayard_2021} uses initial conditions similar to these; however, instead of a superposition of all $N_{\mathrm{exc}}=1$ states, it only includes those in the left half of the system, near the mirror, in an effort to localize energy far from the vacuum. In contrast, this paper attempts to remove any initialization bias. At long times, we expect the most subradiant states to dominate the evolution, regardless of their spatial support. Therefore, we evolve the system for 200 decay times and plot the excited atom population in Fig.~\ref{fig:hwg time evol}. For reference, we also plot the evolution of the system in the absence of interactions. On these timescales, the decay is extremely rapid, whereas the full ordered system decays relatively slowly. This slow-down shows that subradiant states are already populated in the system, but we can further slow the decay using spatial disorder. The position of each atom is randomly offset by some amount $d_i$ such that $z_i = i a + d_i $, where $a$ is the lattice constant of our array. We will sample $d_i$ from a uniform distribution of width $r_d$ such that $d_i \in [-r_d/2,r_d/2]$, and in Fig.~\ref{fig:hwg time evol} we consider maximum disorder $r_d = a = 0.15 \lambda_0$, where $\lambda_0 = 2 \pi c / \omega_0$ is the wavelength of emission. Each curve is averaged over 100 individual trajectories, accounting for both random initial conditions and spatial disorder. Unless otherwise stated, all results shown below use 100 realizations. When averaging different trajectories for $p_{\mathrm{exc}}$, we use the geometric mean throughout this paper. In cases where all trajectories lie within the same order of magnitude, as in Fig.~\ref{fig:hwg time evol}, the geometric and arithmetic mean barely differ. However, in later cases where different trajectories span many different orders of magnitude, the geometric mean offers a more accurate depiction of the average disordered system. Otherwise, very rare, extremely dark states in certain disorder realizations will dominate the late time arithmetic mean. In all other cases besides the excited population, averaging will refer to the arithmetic mean.

We also consider the dynamic fluorescence spectrum of the system as measured from the end of the waveguide, defined as \cite{Rubies_Bigorda_2023,Glauber_1963}
\begin{align}
    S\left(\omega, t^{\prime}\right) = 2 \mathrm{Re}&\Bigg[ \sum_n \int_0^{\infty} d \tau e^{i (k_0 z_n - \omega \tau)}  \nonumber \\ 
    &\times \left\langle\sigma_n^{e g}\left(t^{\prime}+\tau\right) \sigma_n^{g e}\left(t^{\prime}\right)\right\rangle\Bigg] \: .
\end{align}
We evaluate the spectrum at $\gamma_0 t' = 100$ and plot the results in Fig.~\ref{fig:hwg flourescence spectrum}. By this point in the evolution, the noninteracting case has decayed enough that its spectrum is essentially zero. The ordered case displays a clear, slightly shifted resonance, while the disordered case displays two peaks. In addition to slowing decay, we find that disorder has widened the main resonance, similar to the findings in Ref.~\cite{Chen_2022}. Although one typically associates narrower resonances with slower decay, here the widening occurs because many slowly-decaying states with different shifts remain populated at late times.

The behavior in Fig.~\ref{fig:hwg time evol} can be explained using the single-particle spectrum of $H_{\mathrm{hwg}}$. We begin by analyzing the effect of disorder on exchanges between atoms, which we visualize by plotting the spatial support of each eigenmode of the same $N=50$ non-Hermitian Hamiltonian in Fig.~\ref{fig:hwg eigenmodes} for a single realization of spatial disorder. Each column in the figure corresponds to a different eigenmode, and we order the modes with decreasing decay rate. Many of these modes are strongly delocalized, but the slowest decaying ones show clear signs of localization, with support that decays over its nearest neighbors. The fastest decaying modes have most of their support near the edge of the system far from the mirror (bottom left), while the slowly decaying modes have negligible support on this edge and are rather concentrated in the bulk or near the mirror.

In our pursuit of long-lived states, however, we are more interested in the dissipative nature of the Hamiltonian. Thus, we next plot the decay rate of each eigenmode in Fig.~\ref{fig:hwg decay rates}. As a reference, the decay rate of a single, noninteracting emitter is plotted as a horizontal line, such that superradiant modes fall above the line and subradiant below. We then compare the ordered case to the disordered case, averaged over different realizations. Here, disorder keeps the decay rate approximately the same for the modes with large decay rates. As we move to longer-lived states, however, disorder suppresses the decay rates relative to the ordered case. This effect is most pronounced for the slowly-decaying modes at the far right of the spectrum, some of which have decay rates near the numerical precision of our calculations. The late time behavior in Fig.~\ref{fig:hwg time evol} therefore occurs because of these longer-lived modes.

To analyze the effect of the lattice spacing and disorder strength on this system, let's focus our attention on the slowest decaying mode in the spectrum, i.e. $n=N=50$ in our current ordering. This mode will always dominate the late time dynamics for a general initial state and therefore allows us to remove any bias we might include in a choice of initial conditions. In Fig.~\ref{fig:hwg lattice spacing}a, the average decay rate of this slowest mode is plotted as a function of the lattice spacing $a/\lambda_0$ for various disorder strengths $r_d/a$. To avoid any slightly negative decay rates from numerical noise, we take the absolute value and plot $|2 \: \mathrm{Im}\left[ E_N \right]|$. The thinner curves indicate the minimum decay rate achieved over 100 realizations, allowing us to find possible outliers in the average that may dominate the arithmetic mean of the excited population. Beginning with the ordered case $r_d = 0$, we see that as the lattice spacing shrinks, so does the smallest decay rate in the system. In the full waveguide, this occurs because the system approaches the Dicke limit with no coherent interactions and totally symmetric dissipative interactions. In the half waveguide, however, all interactions tend to zero as the atoms bunch up next to the mirror and the two exponentials in $H_{\mathrm{hwg}}$ approximately cancel. At larger lattice spacing, the decay rate fluctuates due to resonances where the mirror causes complete destructive interference and sends all coherent and dissipative interactions to zero. Such resonances occur when $a/\lambda_0 = n/2$ for $n \in \mathbb{Z}$, and two are clearly visible at $a/\lambda_0 = 1$ and $10$ where the decay rate shoots down to zero.

While Ref.~\cite{Fayard_2021} only considered the dilute regime where $a > \lambda_0$, here we further investigate the role of disorder strength over a range of $a$ values. Once we add disorder, the sub-wavelength and super-wavelength cases behave quite differently. For $a > \lambda_0$, we confirm that any strength of disorder will cause localization of energy within the system. As soon as $r_d$ is increased, the decay rate falls sharply to near zero. Here, the slowest mode is localized on about 2 sites but displays no clear phase structure; while fully real, the wavefunction in some cases is purely positive and in other cases has fluctuating sign. This is consistent with the analysis in Ref.~\cite{Fayard_2021}, where it is argued that such states are slowly-decaying because they are localized within the bulk and can't transport their energy to the edge, where decay into the vacuum is non-negligible. The precise phases within such localized states are not crucial to the localization. When we move to the dense regime, where $a < \lambda_0$, we begin to see a large dependence on disorder strength. Weak disorder, where $r_d/a = 0.1$, barely slows decay at all, whereas full disorder pushes the rate down to essentially zero. In this limit, the wavefunctions of the slowest decaying modes show an oscillatory phase structure, where adjacent atoms have opposite sign in $\psi_n$, similar to the asymmetric, subradiant states in the Dicke limit. Because of these two trends, we hypothesize that localization of energy in the bulk and halting transport to the vacuum is no longer the phenomenon behind slowed decay. Rather, the reduction in decay rates seems to come from destructive interference between groups of nearby atoms. As the strength of disorder increases, groups of atoms explore more configurations, and spacings where atoms can destructively interfere become more and more common. Although the presence of the mirror in the half waveguide somewhat muddies this picture, the large dependence on disorder strength and the phase structure of long-lived modes still point to destructive interference as the cause of the reduction in decay rates.

If we now look at the minimum curves, most are around the same order of magnitude as the average curves. The exception to this trend appears for $r_d = 0.5 a$ at small lattice spacing, where the minimum curve is around 5 orders of magnitude lower than the average. Although both curves already have very small decay rates, the difference in scale becomes evident if we evolve the population to very late times. In Fig.~\ref{fig:hwg lattice spacing}b, we evolve to $\gamma_0 t = 10^{11}$ and calculate both the arithmetic and geometric mean once again for $r_d/a = 0.5$ and $a/\lambda_0 = 0.15$. Now, the darkest states across all 100 realizations dominate the arithmetic mean, while the average decay rate determines the geometric mean's path.

We therefore confirm that the half waveguide displays an analog of Anderson localization in the dilute regime and also expand the analysis to different densities and disorder strengths. Once we decrease the lattice spacing, however, the phenomenon behind suppressed decay drastically changes and seems to rely on destructive interference rather than spatial localization, similar to the findings in Ref.~\cite{Skipetrov_2014}.

\section{1D Chain in Free Space}
\label{sec:1d free}

The intuition gained from the waveguide example above can now be translated into the more complex radiative environment of atoms interacting with the electromagnetic continuum of free space. Superradiance in 1D groups of atoms in free space has attracted much research over the past decades. While it was historically studied via scattering in Bose-Einstein condensates \cite{Inouye_1999,Schneble_2003}, experiments have more recently analyzed superradiance in dense clouds of laser-cooled atoms \cite{Ferioli_2021_2,Ferioli_2023}. Such systems also generically host many subradiant modes, so our analysis has experimental relevance in these setups. In free space, we have the full spectrum of modes found in three dimensions, and the master equation is now given as~\cite{Rubies_Bigorda_2023,Lehmberg_1970_1,Lehmberg_1970_2,Asenjo-Garcia_2017}
\begin{align}
    \dot{\rho} = - i \left[ H_{\mathrm{free}} \rho - \rho H_{\mathrm{free}}^{\dagger} \right] + \sum_{ij} \Gamma_{ij} \sigma_i^{ge} \rho \sigma_j^{eg} \: ,
\end{align}
where
\begin{align}
    \label{eq:H free}
    H_{\mathrm{free}} = \omega_0 \sum_{i} \sigma_i^{ee} + \sum_{ij} \left[ - \frac{3 \pi \gamma_0}{\omega_0} \mathbf{d}^{\dagger} \mathbf{G}(\mathbf{r}_{ij},\omega_0) \mathbf{d} \right] \sigma_i^{eg} \sigma_j^{ge} \: ,
\end{align}
and $\Gamma_{ij} = -2 \: \mathrm{Im} \left[ - \frac{3 \pi \gamma_0}{\omega_0} \mathbf{d}^{\dagger} \mathbf{G}(\mathbf{r}_{ij},\omega_0) \mathbf{d} \right]$ are the dissipative interactions. We can also define the coherent interactions as the real part of this expression: $J_{ij} = \mathrm{Re} \left[ - \frac{3 \pi \gamma_0}{\omega_0} \mathbf{d}^{\dagger} \mathbf{G}(\mathbf{r}_{ij},\omega_0) \mathbf{d} \right]$. $\mathbf{d}$ is the transition dipole moment of each atom, and $\mathbf{G}(\mathbf{r},\omega_0)$ is the Green's tensor for a point dipole in a vacuum, given by \cite{Lehmberg_1970_1,Lehmberg_1970_2}
\begin{align}
    G_{\alpha \beta}(\mathbf{r}, \omega) &= \frac{e^{i k r}}{4 \pi r}\Big[\left(1+\frac{i}{k r}-\frac{1}{(k r)^2}\right) \delta_{\alpha \beta} \\ \nonumber
    &+\left(-1-\frac{3 i}{k r}+\frac{3}{(k r)^2}\right) \frac{r_\alpha r_\beta}{r^2}\Big]+\frac{\delta_{\alpha \beta} \delta^{(3)}(\mathbf{r})}{3 k^2} \: .
\end{align}
Here, $\mathbf{r}_{ij} = \mathbf{r}_{i} - \mathbf{r}_{j}$ is the displacement between atoms $i$ and $j$ and $k = \omega/c$. The diagonal dissipative interactions are given by the individual atoms' free space decay rate $\Gamma_{ii} = \gamma_0$. We also absorb the Lamb shifts $J_{ii}$ into the definition of $\omega_0$. In 1D and 2D, we consider $\mathbf{d}$ perpendicular to the array, while in 3D we consider it oriented along one of the cube axes.

\begin{figure}
	\centering
	
	\begin{subfigure}{\linewidth}
		\caption{}
		\includegraphics[width=8.6cm]{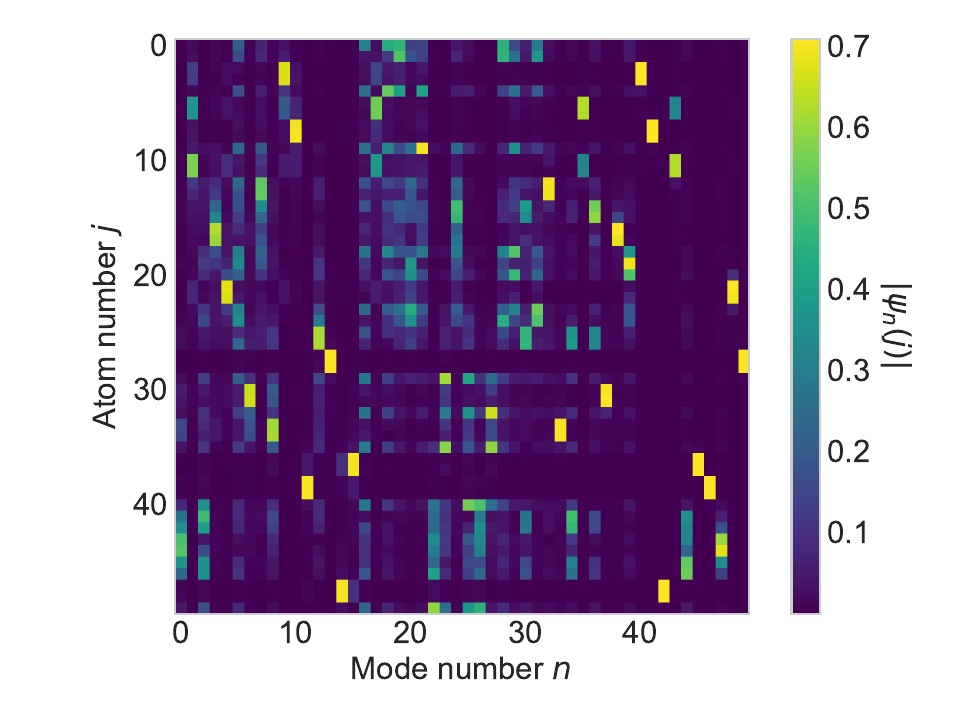}
		\label{fig:1d free eigenmodes}
	\end{subfigure}
	
	\begin{subfigure}{\linewidth}
		\caption{}
		\includegraphics[width=8.6cm]{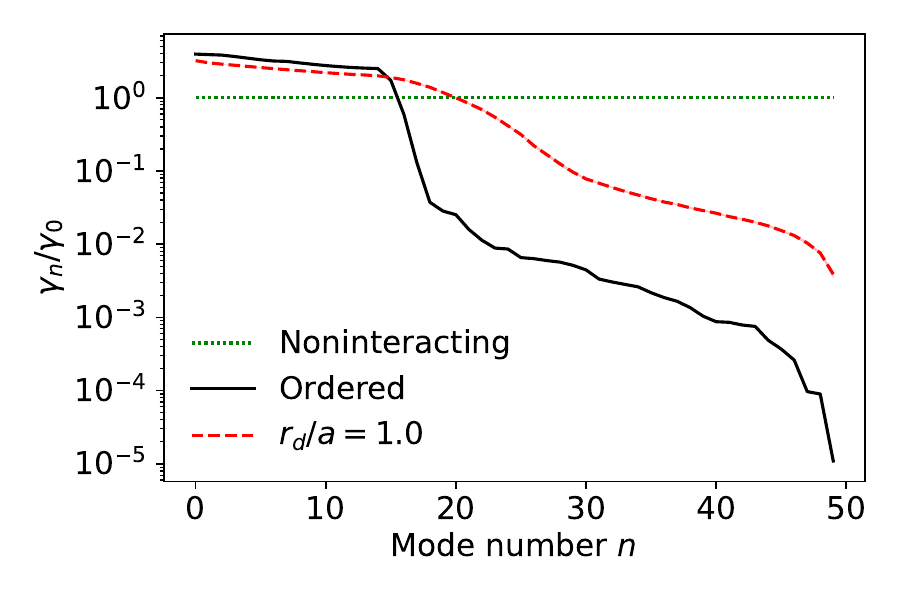}
		\label{fig:1d free decay rates}
	\end{subfigure}

	\caption{The spatial profile $|\psi_n(j)|$ (a) and decay rates $\gamma_n = -2 \: \mathrm{Im}[E_n]$ (b) of single-excitation eigenmodes of the non-Hermitian Hamiltonian of an $N=50$ 1D chain in free space. All parameters besides the photon modes are the same as in Fig.~\ref{fig:hwg modes}.}
	\label{fig:1d free modes}
\end{figure}

\begin{figure}
	\centering
	\includegraphics[width=8.6cm]{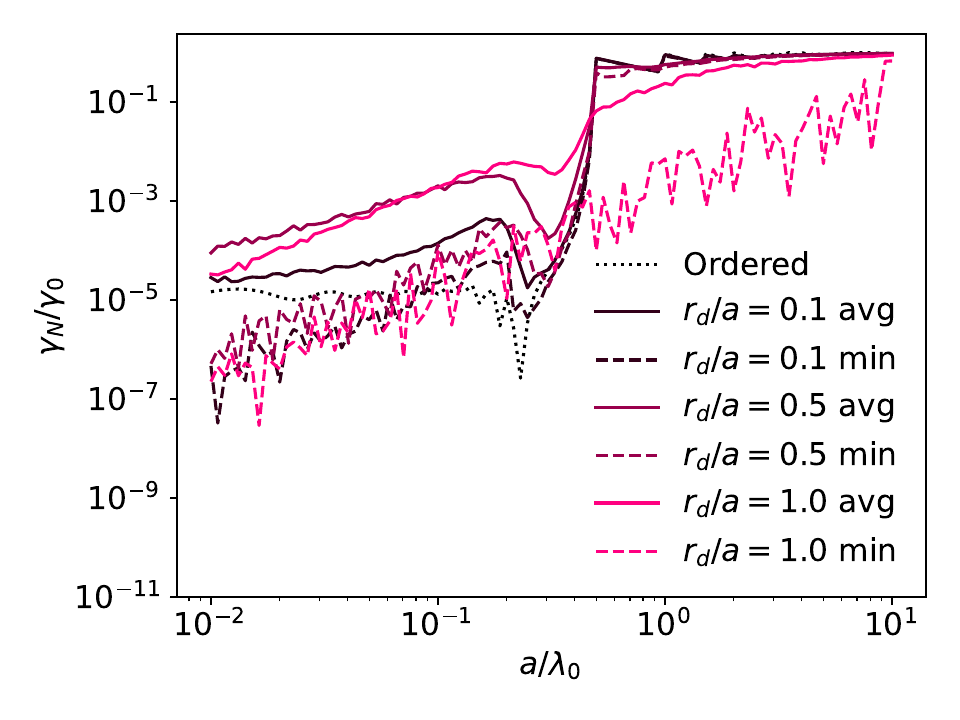}
	\caption{Magnitude of the slowest decay rate $\gamma_N$ in the spectrum of an $N=50$ 1D chain in free space as a function of lattice spacing and disorder strength. The minima curves show very dark states appearing in certain realizations.}
	\label{fig:1d free lattice spacing}
\end{figure}

We now analyze the effect of spatial disorder on a one-dimensional chain with $N=50$ atoms in free space with the same parameters as the half waveguide. Such a chain is pictured in Fig.~\ref{fig:schematics}b. The eigenmodes of the single excitation manifold are plotted in Fig.~\ref{fig:1d free eigenmodes}. Compared to the half waveguide, there are now many localized states across the spectrum, with many being concentrated on a single pair of atoms. Despite this, we don't expect the system to show any open system analog of localization because of disorder's effect on decay rates. These are plotted in Fig.~\ref{fig:1d free decay rates}, analogous to Fig.~\ref{fig:hwg decay rates}. Now, disorder increases the decay rate not just for a majority of eigenmodes but also to the long-lived, subradiant modes that originally persisted at long time scales. Thus, although disorder spatially localizes eigenmodes, it accelerates the system's dissipation.

We can gain more insight into this effect by considering again the slowest decaying mode in the spectrum and how its decay rate depends on lattice spacing and disorder strength. We plot these curves in Fig.~\ref{fig:1d free lattice spacing}a in direct analogy to Fig.~\ref{fig:hwg lattice spacing}a. We now observe a sharp phase transition from approximately noninteracting emission to subradiance as we decrease the lattice spacing in the ordered case. Ref.~\cite{Henriet_2019} explains how, as $a$ decreases past $\lambda_0/2$, certain states can no longer decay transverse to the chain and must propagate longitudinally through the entire array to escape to the vacuum. These states therefore become very subradiant and cause the transition observed in the figure. If we then increase the disorder strength, we slowly wash out the transition and find a smoother, approximately power law trend between the slowest decay rate and the lattice spacing. Examining the phase structure of the slowest decaying states in all disorder realizations reveals they are two-body singlet states between the most closely-packed atoms. Spatial disorder will, by chance, push two atoms very close together; in this limit, their decay rates and eigenstates will match those of a Dicke superradiant system, and will therefore form a two-body dark state given by a singlet (these pairs will also create much brighter symmetric triplet states, which appear in the spectrum around $n = 15$ in Fig.~\ref{fig:1d free eigenmodes}). The trends in Fig.~\ref{fig:1d free lattice spacing} can therefore be understood as a comparison between the decay rate of the darkest many-body eigenstate and the average decay rate of few-body dark states of closely packed atoms.

At large $a$, where the ordered case hasn't transitioned to the subradiant regime yet, we see that increasing disorder reduces the decay rate. Since the many-body state still decays at approximately the vacuum rate $\gamma_0$, increasing the decay allows more approximate two-body singlets to form, creating more dark Dicke states and lowering the decay rate. The darkest of these states, the closely-packed singlets, can be seen in the minimum $r_d/a=1$ curve, which reaches many orders of magnitude lower than both the ordered case and the average rate. The weaker disorder cannot create such close-packed pairs, and therefore remain close to their averages. However, at small $a$, we see a dramatically different trend. The many-body state is now fully subradiant, and the destructive interference which causes this relies on the regular spacing of the lattice. As such, we expect disorder to actually increase the decay rate, as it destroys the system-wide destructive interference. The many-body approximate Dicke state has a smaller decay rate than the two-body approximate Dicke state, and therefore no advantage is gained with disorder. We also see that in the extremely small $a/\lambda_0$ limit, the largest decay rate is achieved by $r_d/a = 0.5$. While $r_d/a = 0.1$ has not fully destroyed the destructive interference of the ordered case and $r_d/a = 1$ can create extremely close-packed pairs of atoms, the intermediate disorder case has neither of these advantages and amplifies decay the most. In contrast, all disorder strengths have essentially overlapping minimum curves at small $a$, and at small enough lattice spacing, they overtake the ordered case and suppress decay. At such densities, all disorder strengths can create close-packed pairs, and as they get closer and closer, the darkest among them can overtake the many-body subradiant state. 

Finally, we note the dip in decay rates around $a/\lambda_0 \approx 0.2$. This dip was also observed in Ref.~\cite{Rubies_Bigorda_2023} as a peak in subradiant population and was attributed to a balance between the approach to completely dark states in the Dicke limit and the coherent transfer between subradiant and superradiant states. We note that the exact value of $a/\lambda_0$ where this occurs is different between our analysis and that of Ref.~\cite{Rubies_Bigorda_2023} because we consider the most subradiant single-excitation eigenstate in contrast to a general multi-excitation state.

\section{Higher dimensional arrays}
\label{sec:higher d arrays}

\begin{figure*}
	\centering
	
	\begin{subfigure}{0.49\linewidth}
		\caption{}
        \includegraphics[width=8.6cm]{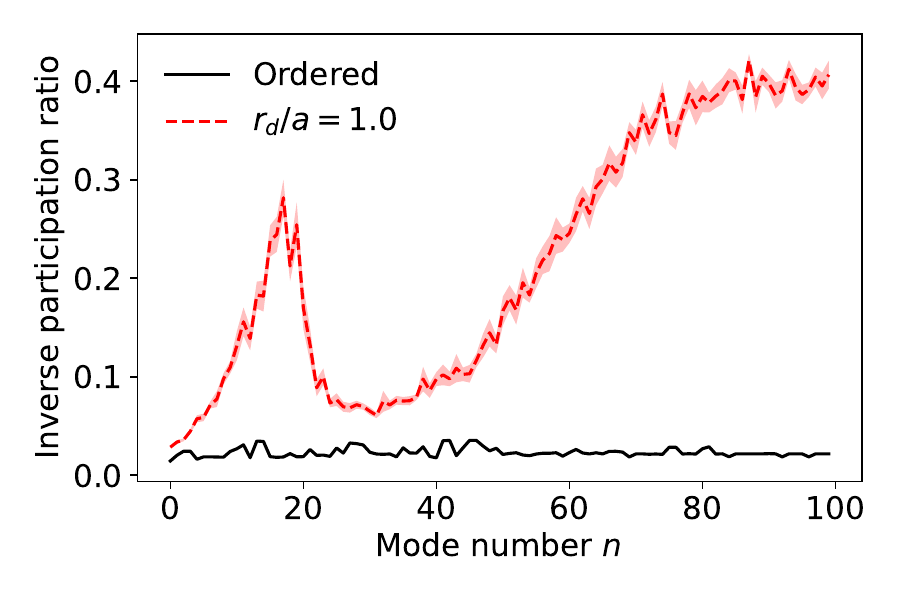}
		\label{fig:2d free iprs}
	\end{subfigure}\hfill
    \begin{subfigure}{0.49\linewidth}
		\caption{}
        \includegraphics[width=8.6cm]{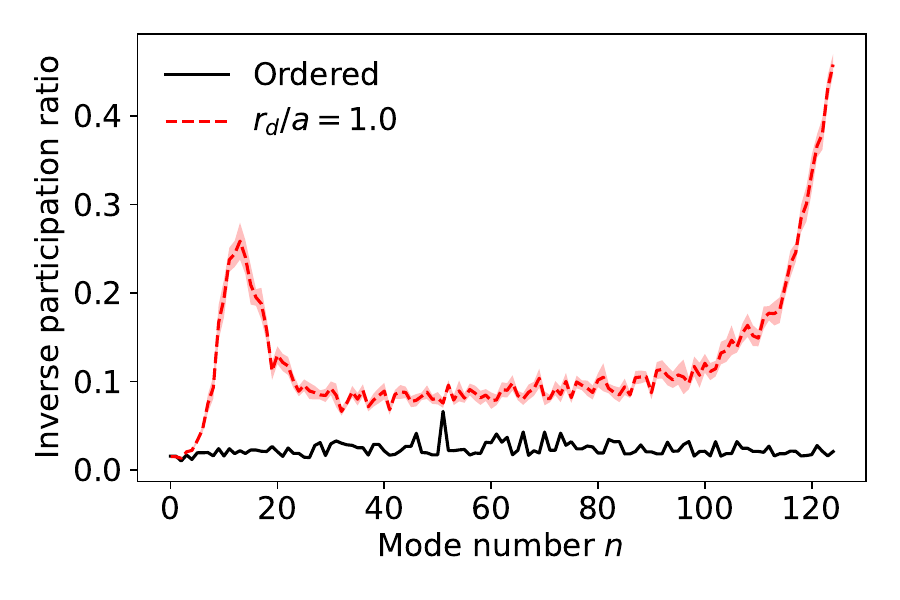}
		\label{fig:3d free iprs}
	\end{subfigure}

    \begin{subfigure}{0.49\linewidth}
		\caption{}
        \includegraphics[width=8.6cm]{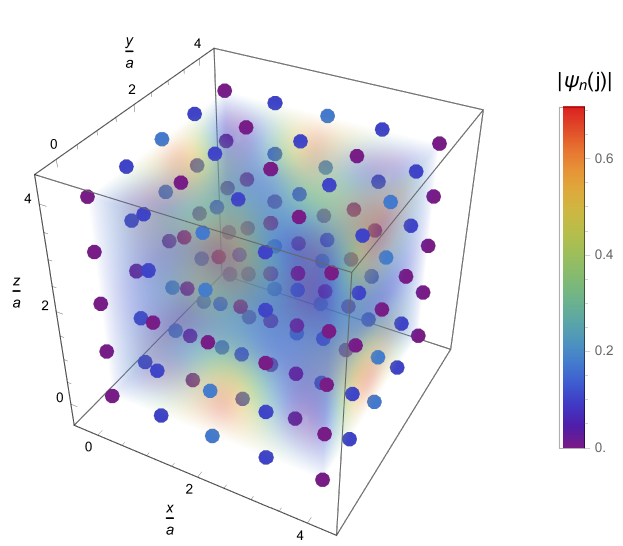}
		\label{fig:ordered555}
	\end{subfigure} \hfill
	\begin{subfigure}{0.49\linewidth}
		\caption{}
        \includegraphics[width=8.6cm]{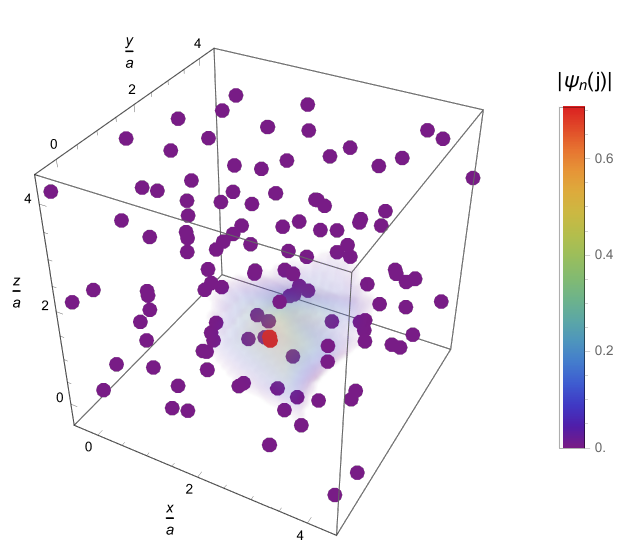}
		\label{fig:disordered555}
	\end{subfigure}

	\caption{The inverse participation ratios of single-excitation eigenmodes of the non-Hermitian Hamiltonian of an $N=10\times10$ 2D square (a) and an $N=5\times5\times5$ 3D cube (b) in free space. We also visualize the spatial support of the slowest decaying mode in both an ordered (c) and disordered (d) $5\times5\times5$ array of atoms. In the ordered case, the mode has support across the system with slightly more participation at the edges. In the disordered case, all but two atoms are essentially unoccupied, and the two occupied atoms are so close together they appear on top of each other. Analysis of the wavefunction reveals the atoms are in an approximate singlet state, the dark state from the two-atom Dicke limit.}
	\label{fig:23d free modes}
\end{figure*}

\begin{figure*}
    \centering

    \begin{subfigure}{0.49\linewidth}
		\caption{}
        \includegraphics[width=8.6cm]{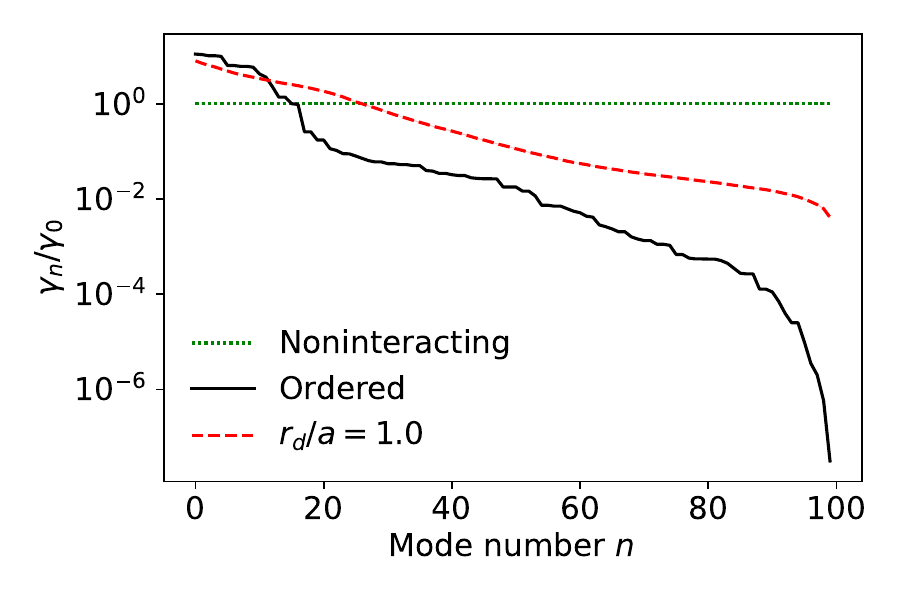}
		\label{fig:2d free decay rates}
	\end{subfigure}\hfill
    \begin{subfigure}{0.49\linewidth}
		\caption{}
        \includegraphics[width=8.6cm]{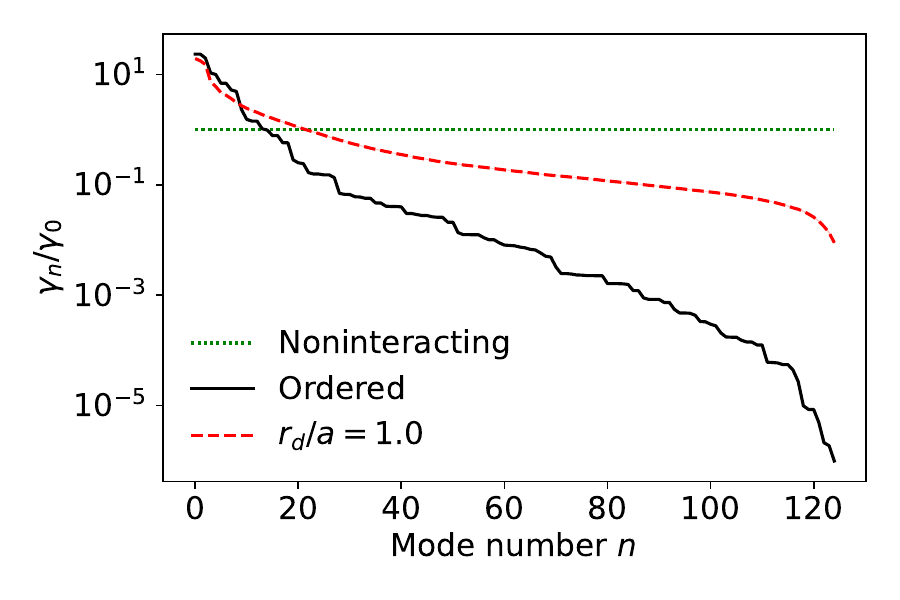}
		\label{fig:3d free decay rates}
	\end{subfigure}

    \caption{Decay rates $\gamma_n = -2 \: \mathrm{Im}[E_n]$ of single-excitation eigenmodes of the non-Hermitian Hamiltonian of an $N=10\times10$ 2D square (a) and an $N=5\times5\times5$ 3D cube (b) in free space.}
\end{figure*}

\begin{figure*}
	\centering
	
	\begin{subfigure}{0.49\linewidth}
		\caption{}
        \includegraphics[width=8.6cm]{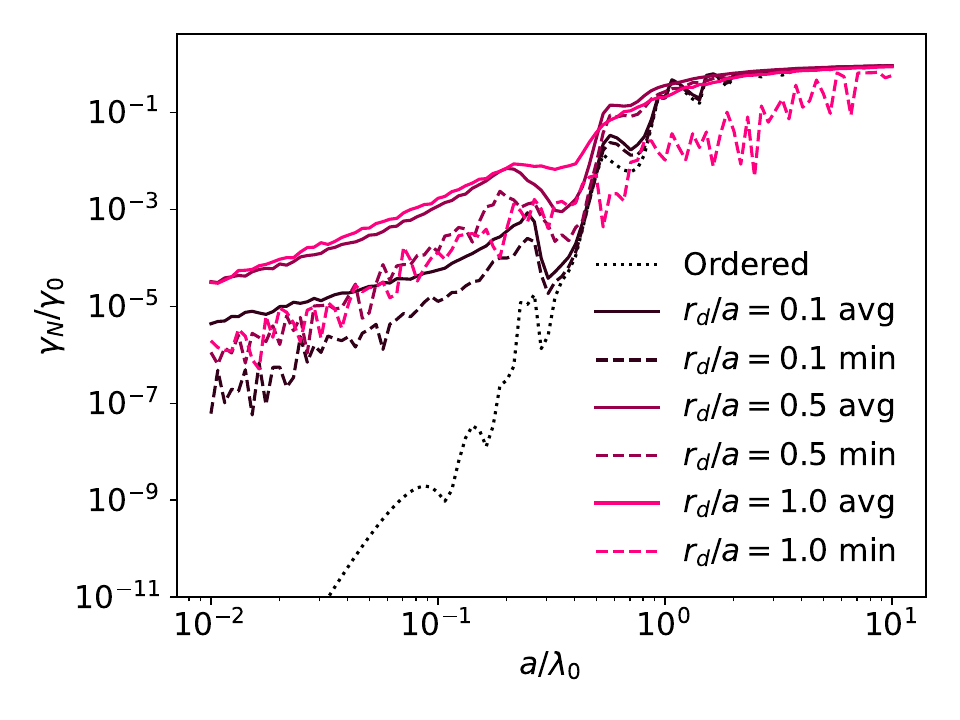}
		\label{fig:2d free lattice spacing}
	\end{subfigure}\hfill
    \begin{subfigure}{0.49\linewidth}
		\caption{}
        \includegraphics[width=8.6cm]{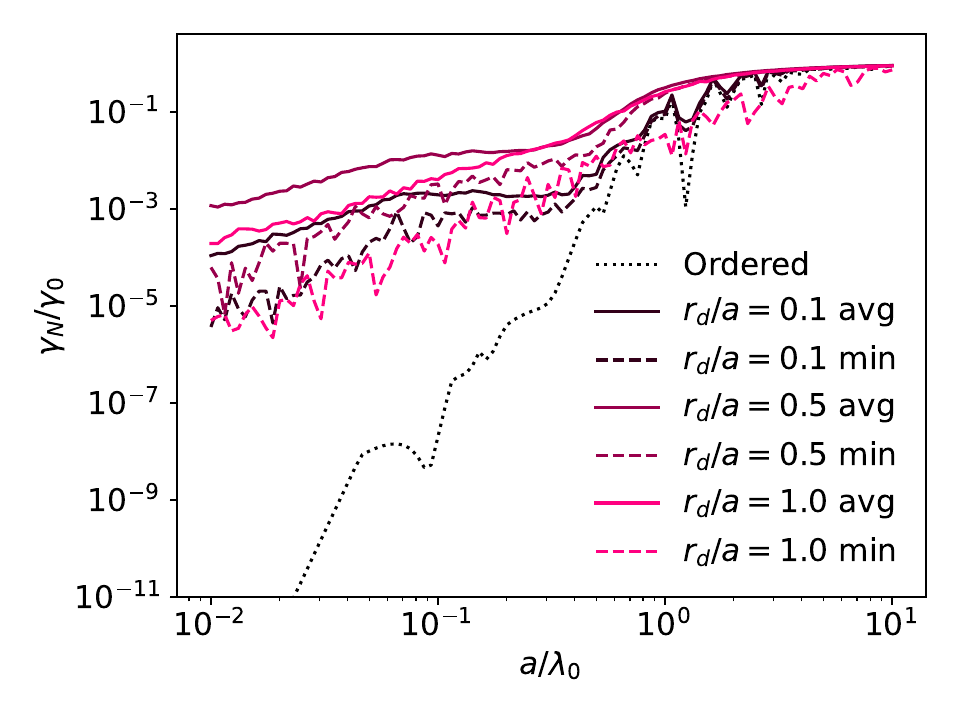}
		\label{fig:3d free lattice spacing}
	\end{subfigure}

	\caption{Magnitude of the slowest decay rate $\gamma_N$ in the spectrum of an $N=10\times10$ 2D square (a) and an $N=5\times5\times5$ 3D cube in free space as a function of lattice spacing and disorder strength.}
	\label{fig:23d free lattice spacing}
\end{figure*}

Given this picture, we now move to 2D and 3D arrays of atoms in free space as pictured in Figs.~\ref{fig:schematics}c-d. As it is less intuitive to plot the support of eigenmodes on a single axis now, we will opt for plotting the inverse participation ratio (IPR) of each mode instead, defined as $\mathrm{IPR}(n) = \sum_{j}|\psi_n(j)|^4$. For states localized on a single site, $\mathrm{IPR} = 1$, while completely delocalized states have $\mathrm{IPR} = 1/N$.

In Fig.~\ref{fig:2d free iprs}, we consider a $N=10\times10$ square array and calculate the IPR for both the ordered and maximally disordered cases, the latter being averaged over the standard 100 realizations. We see that all disordered modes are more localized than the ordered case, and in general the slower decaying modes ($n \rightarrow N = 100$) are more localized. The exception to this trend appears around $n \lesssim 20$, where a large peak in the IPR appears. Just as in the 1D case, this can be attributed to the bright, approximately symmetric triplet states created between closely-packed pairs of atoms, orthogonal to the dark two-body singlet states appearing on the right of the graph. Despite spatial localization, we again find that most decay rates are amplified, as shown in Fig.~\ref{fig:2d free decay rates}. In Figs.~\ref{fig:3d free iprs} and \ref{fig:3d free decay rates}, we consider a $N = 5\times5\times5$ 3D cubic array, and the same observations apply. We also visualize the most subradiant modes in 3D in both the ordered and disordered cases in Figs.~\ref{fig:ordered555},\ref{fig:disordered555}. Just as predicted, the disordered case is limited to the two closest atoms in the cloud.

If we again focus on the most subradiant mode's decay rate in Figs.~\ref{fig:2d free lattice spacing},\ref{fig:3d free lattice spacing}, we find that there is no sharp transition between plateaus in 2D and 3D. Once the array becomes sub-wavelength, the ordered decay rate diminishes without any sort of plateau, and by the time $a/\lambda_0 =  0.01$, the slowest decay rate is about 6 orders of magnitude smaller than in the 1D case. Below the figure's border, numerical noise causes unphysical fluctuations in the decay rate, so we leave out these data points. The curves for spatial disorder, on the other hand, follow roughly the same trend as they do in Fig.~\ref{fig:1d free lattice spacing}. Looking to the dilute regime in 2D, we see a couple of choices for $a$ where disorder slows decay. However, this suppression of decay is not as strong as in 1D and does not persist for all values of $a \gtrsim \lambda_0$; only the minimum $r_d/a = 1.0$ curve shows any substantial suppression of decay. By the time we get to 3D, any advantage from disorder seems to be washed away for the average disorder realization, while the arithmetic mean will still be dominated by dark outliers. Taken together, these observations seem to support the two- and few-body dark state picture, as the higher dimensionality in 2D and 3D should make close few-body systems rarer and therefore make disorder slightly less effective at slowing decay. On the other hand, the higher connectivity in higher dimensions clearly allows the ordered case to create longer-lived subradiant states. Thus, as we move to higher and higher dimensions, we expect many-body subradiant states to always decay more slowly than the disordered few-body subradiant states.

While the examples shown here were shown for a $10\times 10$ square and a $5\times 5 \times 5$ cube, a more detailed analysis on the $N$-dependence of the subradiant transition can be found in Appendix~\ref{sec:transition scaling}.

\section{Conclusion}
\label{sec:conc}

We analyzed the effect of disorder on subradiant single-excitation states in multiple different atom array configurations. Motivated by previous work in 1D waveguides and 3D free space clouds, we searched for analogs of Anderson localization to create extremely long-lived states using spatial disorder. Although disorder always slows decay in the half waveguide, disorder in free space can either speed up or slow down decay depending on the relative lattice spacing of the array. In the dilute half waveguide, eigenstates are localized and display no particular phase structure, and any strength of disorder will cause dramatic suppression of decay. However, as the lattice spacing is decreased, the phase of the longest-lived states begins to oscillate on adjacent sites, and the relative suppression of decay depends strongly on the strength of disorder. These trends seem to imply that while the dilute half waveguide slows decay because of a true localization of energy within the bulk, the dense case relies on destructive interference within groups of atoms packed close together by strong disorder.

In free space, the effect of spatial disorder can also be understood through its creation of close-packed few-body subradiant states. This effect is primarily seen through the creation of pairs of atoms spaced very close to each other, effectively making other atoms irrelevant. These two atoms then form an approximate Dicke system, with a bright symmetric triplet state and a very dark antisymmetric singlet state. Whether disorder slows down decay in free space then becomes a question of how this average two-body dark state compares to the ordered, many-body dark state. In 1D, we observe a set of lattice spacings $a \gtrsim \lambda_0$ where disorder can slow decay, but as we move to 2D and 3D, this effect is washed away.

We observe spectral localization in all of our setups, but, in contrast with closed systems, this does not generally guarantee a dynamical localization of energy within the array. Although disorder can sometimes slow decay, our analysis reveals that it relies on the creation of few-body dark states and does not persist in higher dimensions. Although the average over these disordered geometries doesn't generally slow decay, one could engineer the spacing of atoms within the chain to create certain bright and dark states, allowing the modification of cooperative decay.

Our analysis opens up multiple questions for future study. One of the most obvious next steps is to consider states with more than one excitation. Although this makes it harder to analyze the eigenmodes of the Hamiltonian, and although $N_{\mathrm{exc}}=1$ states are generically the most subradiant, the possibility of an analog of MBL in open arrays warrants study. Another next step would involve an in-depth analysis of the effect of dynamic disorder as opposed to static shifts in position. Although one could imagine having each atom do a random walk within its unit cell, the exact implementation of such disorder is up for debate. The eigenmode analysis also would not apply, and one would instead have to rely on numerical evolution or perturbation theory as in Ref.~\cite{Chen_2022}. Finally, one can consider other interactions besides waveguides and free space dipole-dipole interactions. One such possibility is the soft-core interactions in a Rydberg dressed system \cite{Young_2023,Pupillo_2010,Johnson_2010,Gil_2014}. In this setup, one slightly mixes the excited state with a Rydberg state using an off-resonant driving. The Rydberg blockade then creates a soft-core interaction at small distances, while at large distances it resembles a van der Waals interaction. Since this interaction is multi-excitation, it does not affect the single-excitation dynamics studied here, but future works with $N_{\mathrm{exc}}>1$ could analyze the effect of such dressing.

\section{Acknowledgements}

We would like to thank Oriol Rubies-Bigorda, Hanzhen Ma, and Yidan Wang for fruitful discussions. We also thank Harvard University's FAS Research Computing for numerical resources. N.O.G. acknowledges support from the Harvard Purcell and Generation Q G2 fellowships; S.O. is supported by a postdoctoral fellowship of the Max Planck Harvard Research Center for Quantum Optics. S.F.Y. acknowledges NSF via PHY-2207972 and the CUA PFC and AFOSR via Grant No. FA9550-19-1-0233.


%

\appendix

\section{$N$-Dependence of Subradiant Transition}
\label{sec:transition scaling}

\begin{figure*}
	\centering
	
	\begin{subfigure}{0.33\linewidth}
		\caption{}
        \includegraphics[width=5.4cm]{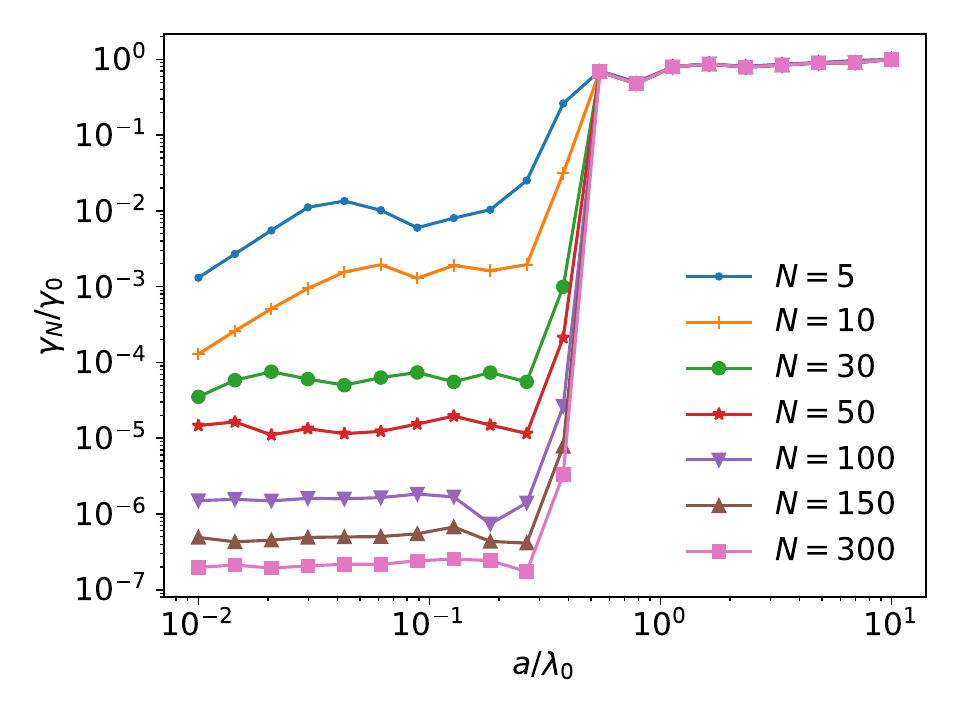}
		\label{fig:1d N scaling}
	\end{subfigure}\hfill
    \begin{subfigure}{0.33\linewidth}
		\caption{}
        \includegraphics[width=5.4cm]{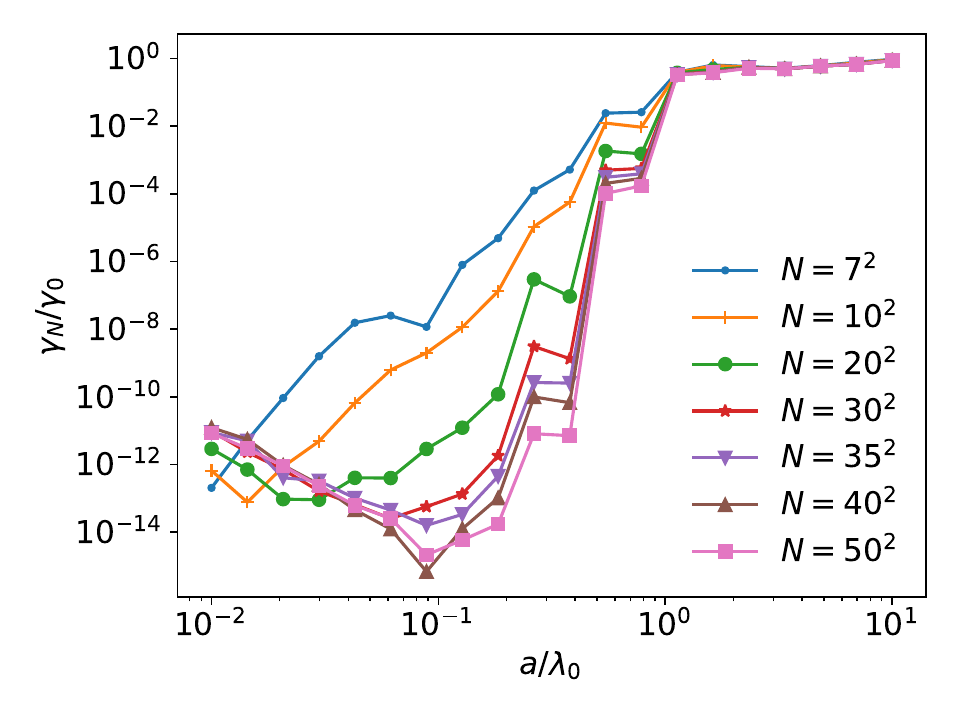}
		\label{fig:2d N scaling}
	\end{subfigure}\hfill
    \begin{subfigure}{0.33\linewidth}
		\caption{}
        \includegraphics[width=5.4cm]{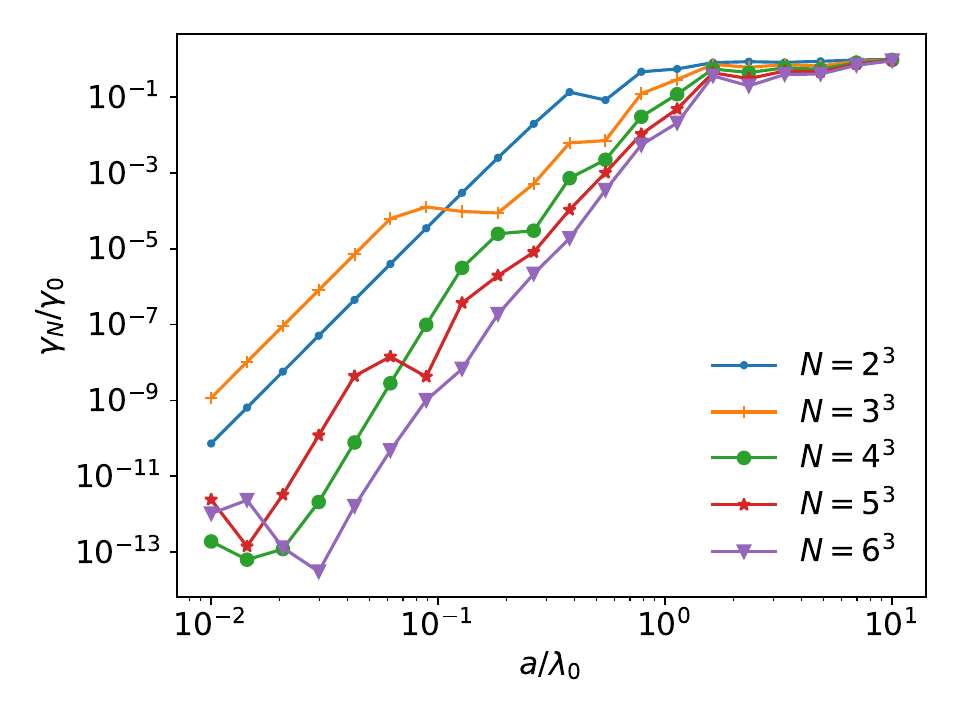}
		\label{fig:3d N scaling}
	\end{subfigure}

\caption{Scaling of the transition to subradiance in the slowest decaying mode as we increase $N$ in free space with no disorder. In 1D (a), we see the transition get sharper and sharper as the number of atoms increases. In 2D (b) and 3D (c), while the shift to subradiance does get sharper with increasing $N$, the crossover does not occur at one value of $a/\lambda_0$.}
	\label{fig:N scaling}
\end{figure*}

To analyze the transition observed in the slowest decaying mode from normal decay to subradiance, we consider many different arrays of varying size $N$ in Fig.~\ref{fig:N scaling}. Restricting ourselves to the ordered case, we consider 1D (a), 2D (b), and 3D (c) systems and calculate this slowest decay rate in the spectrum as a function of lattice spacing. In 1D, we observe the previously mentioned transition getting sharper and sharper. It also occurs at the single value of $a/\lambda_0 = 0.5$ for all values of $N$, as mentioned in Ref.~\cite{Henriet_2019}. In 2D and 3D, however, this picture breaks down, and we instead see a crossover into subradiance. Although increasing $N$ in higher dimensions does amplify subradiance and make the change sharper, there is still no single value of $a/\lambda_0$ where we observe a transition. This lack of a sharp transition is one reason why disorder in systems with $a \gtrsim \lambda_0$ does not provide a subradiant advantage as it does in 1D.

\section{Entanglement Evolution}
\label{sec:entanglement}

\begin{figure*}
	\centering
	
	\begin{subfigure}{0.33\linewidth}
		\caption{}
        \includegraphics[width=5.4cm]{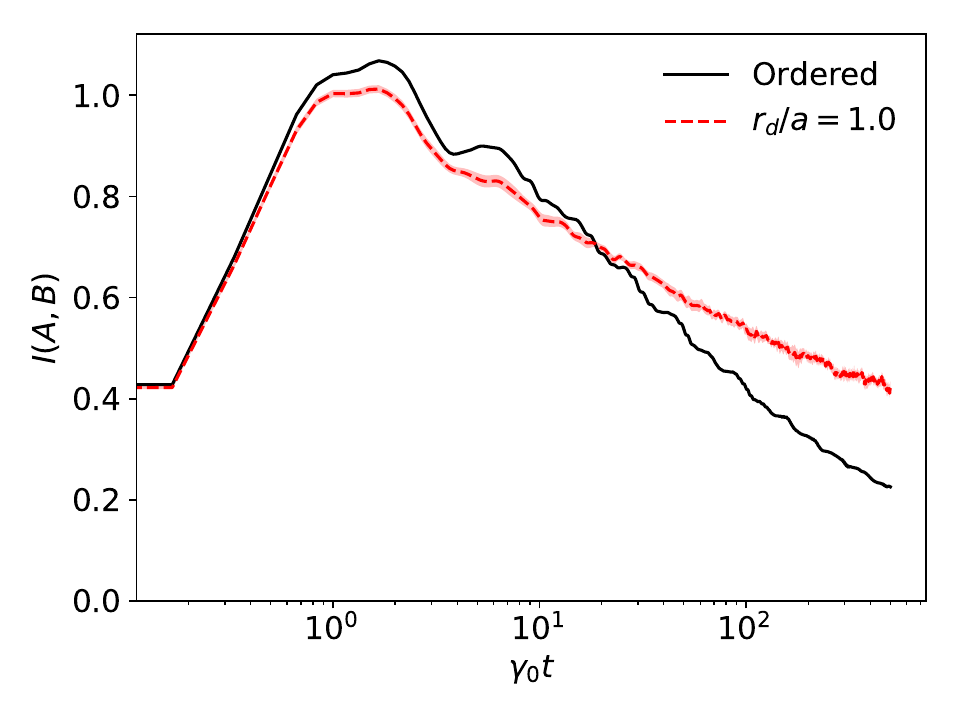}
		\label{fig:hwg mutual info}
	\end{subfigure}\hfill
    \begin{subfigure}{0.33\linewidth}
		\caption{}
        \includegraphics[width=5.4cm]{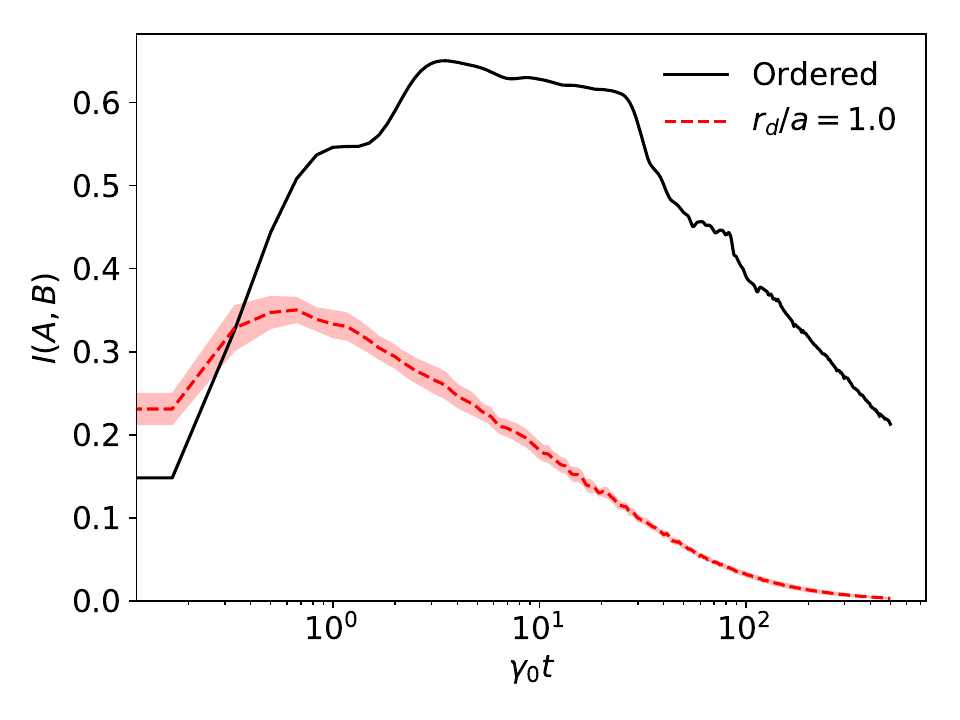}
		\label{fig:1d free mutual info dense}
	\end{subfigure}\hfill
    \begin{subfigure}{0.33\linewidth}
		\caption{}
        \includegraphics[width=5.4cm]{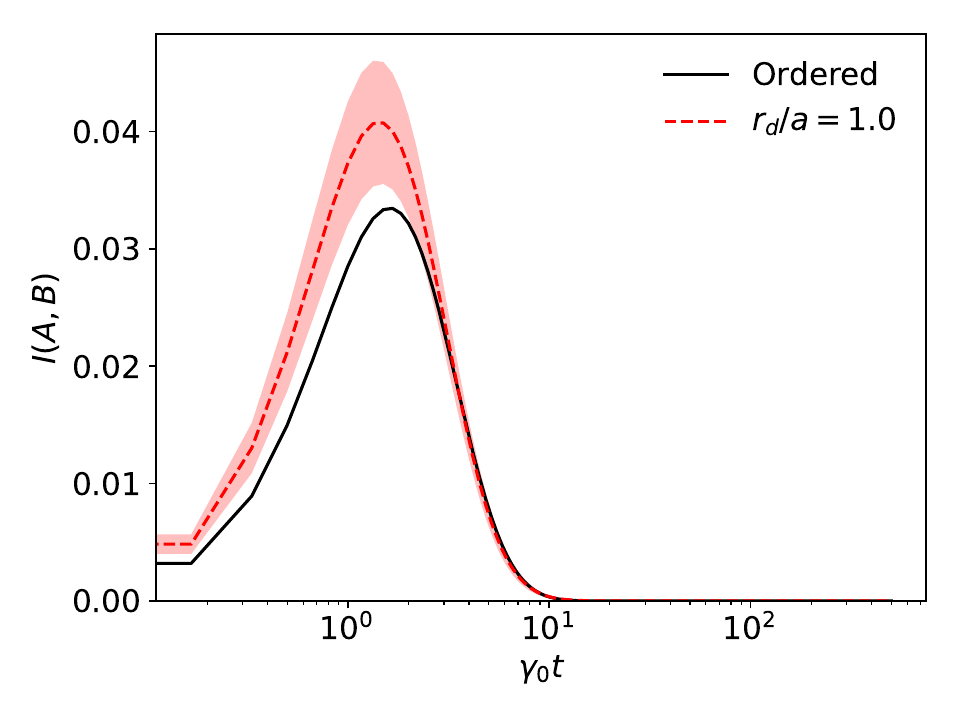}
		\label{fig:1d free mutual info dilute}
	\end{subfigure}

	\caption{Mutual information between two halves of a $N=50$ chain in the half waveguide with $a/\lambda_0 = 0.15$ (a), free space with $a/\lambda_0 = 0.15$ (b), and free space with $a/\lambda_0 = 1.35$ (c). The system is initialized with the $j=26$th atom excited and evolves both without disorder and with maximum disorder.}
	\label{fig:mutual info evol}
\end{figure*}

Recent work has examined measures of entanglement, like average pairwise concurrence \cite{Santos_2022}, in characterizing subradiant modes in dissipative systems. We calculate the entanglement generated during our simulations as an additional axis to analyze the presence or absence of localization. Spatial localization should slow the initial growth of entanglement in the system, just as in traditional Anderson or many body localization. In addition, a slowed decay should extend the lifetime of said entanglement before the system relaxes to the entanglement-free ground state. Thus, our analog of localization would correspond to a broader, shorter peak in entanglement compared to the ordered case. Instead of average pairwise concurrence, we use the mutual information between two halves of the system $A$ and $B$, given by
\begin{align}
    I(A,B) = S(A) + S(B) - S(A,B) \: ,
\end{align}
where $S$ is the von Neumann entropy of a given subsystem
\begin{align}
    S(A) = - \Tr \left[ \rho_A \log \rho_A \right]  .
\end{align}
$S(A,B)$ refers to the von Neumann entropy of the entire system's mixed state. In Fig.~\ref{fig:hwg mutual info}, the mutual information of the fully disordered $N=50$ half waveguide from Fig.~\ref{fig:hwg modes} is plotted. Rather than a randomized initial state, we choose the product state with only the atom at $j = 26$ excited. The mutual information then tracks both the spreading and lifetime of this excitation. In the ordered case, the quick build-up of coherence typically associated with collective decay causes an initial peak in the mutual information. However, as the excitation decays, this entanglement is slowly lost. As expected, the presence of disorder slows the spreading of mutual information, resulting in a lower peak, and extends the lifetime of the excitation, resulting in a longer, higher tail. We note that choosing a larger value for $a$ does not qualitatively affect the graph, as even for large spacing the system has slowly-decaying localized states.

Turning to free space, we repeat the calculations for the fully disordered $N=50$ 1D chain from Fig.~\ref{fig:1d free modes} (with $a/\lambda_0 = 0.15$) and plot the results in Fig.~\ref{fig:1d free mutual info dense}. Although the first time step of the disordered curve has a large average mutual information than the ordered case, the peak is still lower and the initial rate of growth is also shallower. The accelerated decay mentioned in Sec.~\ref{sec:1d free} also causes the tail to decay much faster than in the ordered case. In this case, disorder suppresses entanglement at both early and late times.

Finally, we consider a dilute 1D chain in free space, with $a/\lambda_0 = 1.35$, in Fig.~\ref{fig:1d free mutual info dilute}. Now, we still have spectral localization, but we also slow decay on average with disorder. Because the ordered case has a larger decay rate in the dilute regime, the peak in mutual information is much smaller, but because disorder suppresses this decay, it actually reaches a larger maximum mutual information before dropping near zero. Although it's not visible on these scales, the slow-down in decay also extends the tail of the disordered curve compared to the ordered one. Although spectral localization and decay suppression were both visible in the half waveguide, we see that here the shift in decay rates dominates the behavior of the mutual information.

\section{On-Site Detuning Disorder}
\label{sec:freq dis}

\begin{figure*}
	\centering
	
	\begin{subfigure}{0.49\linewidth}
        \caption{}
		\includegraphics[width=8.6cm]{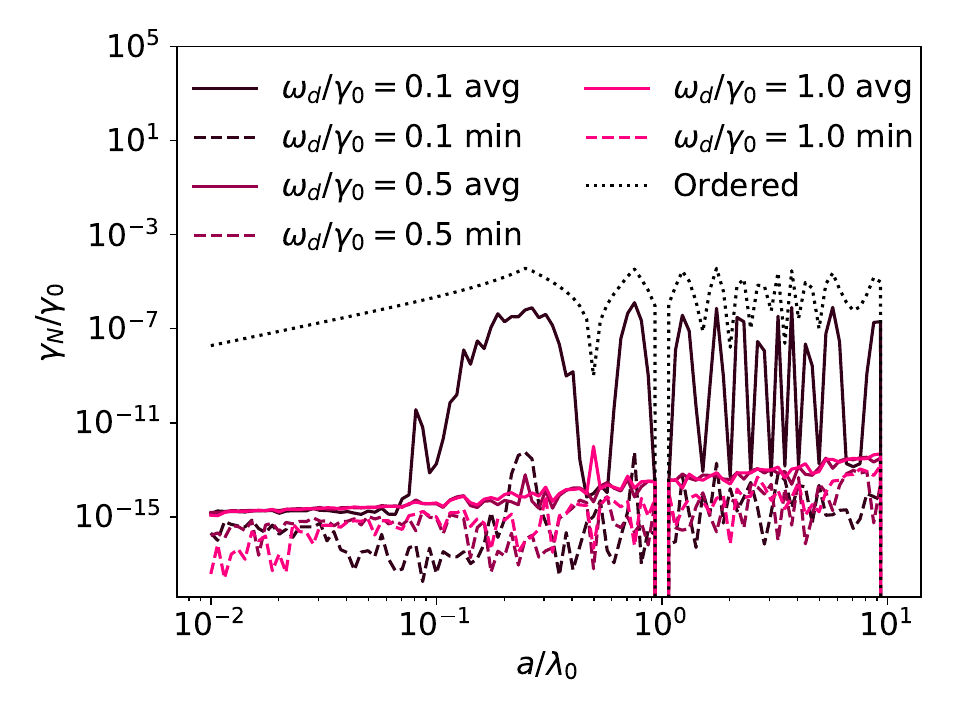}
		\label{fig:hwg detuning}
	\end{subfigure}\hfill
    \begin{subfigure}{0.49\linewidth}
		\caption{}
        \includegraphics[width=8.6cm]{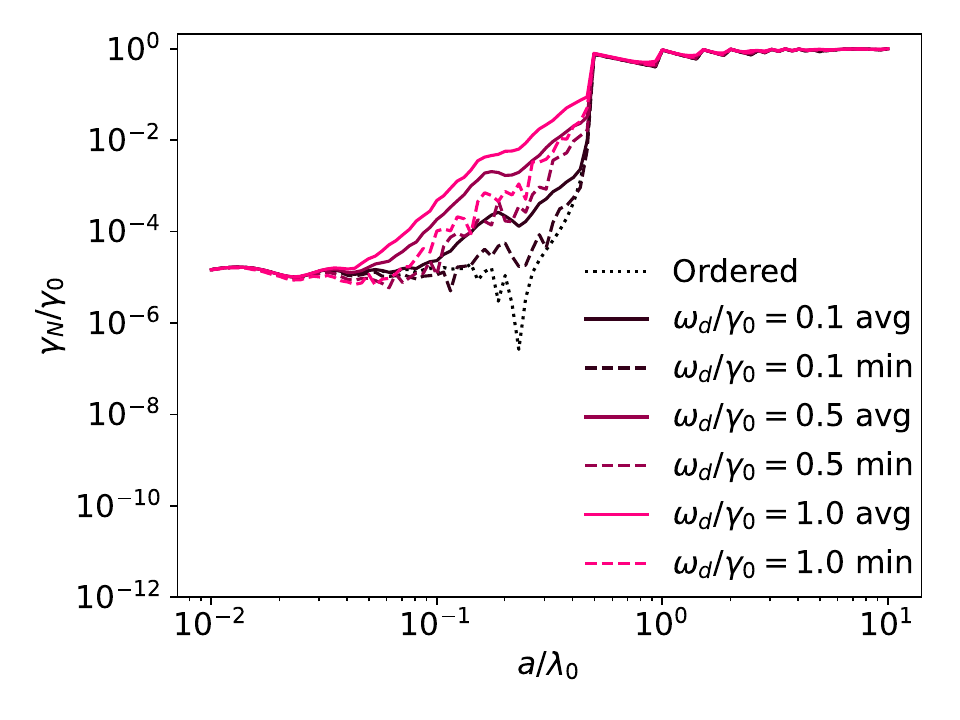}
		\label{fig:1d free detuning}
	\end{subfigure}

    \begin{subfigure}{0.49\linewidth}
        \caption{}
		\includegraphics[width=8.6cm]{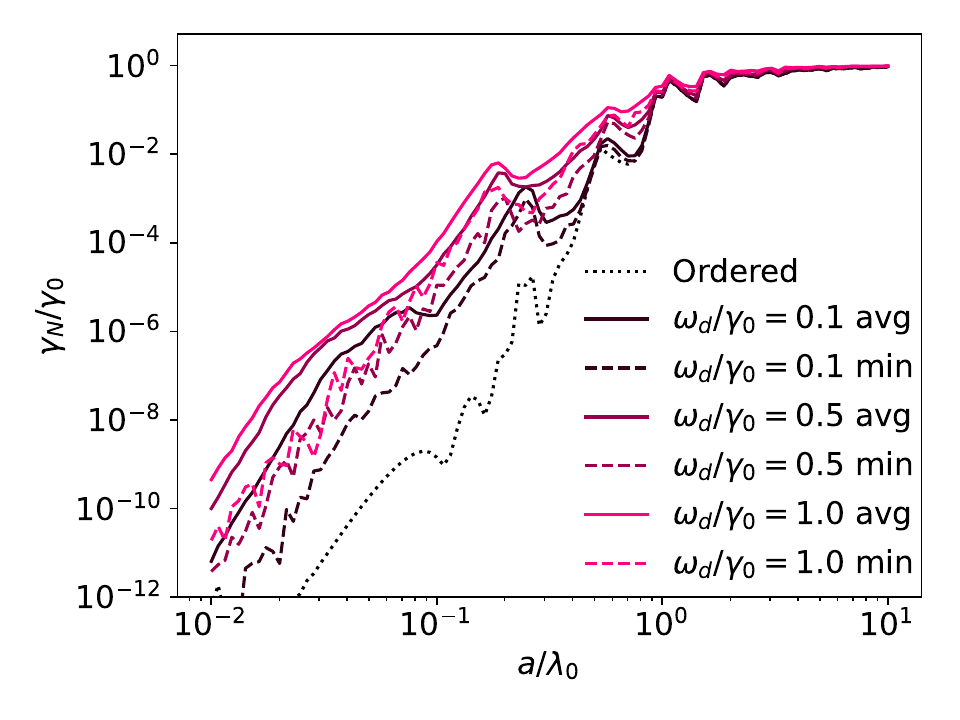}
		\label{fig:2d free detuning}
	\end{subfigure} \hfill
	\begin{subfigure}{0.49\linewidth}
        \caption{}
		\includegraphics[width=8.6cm]{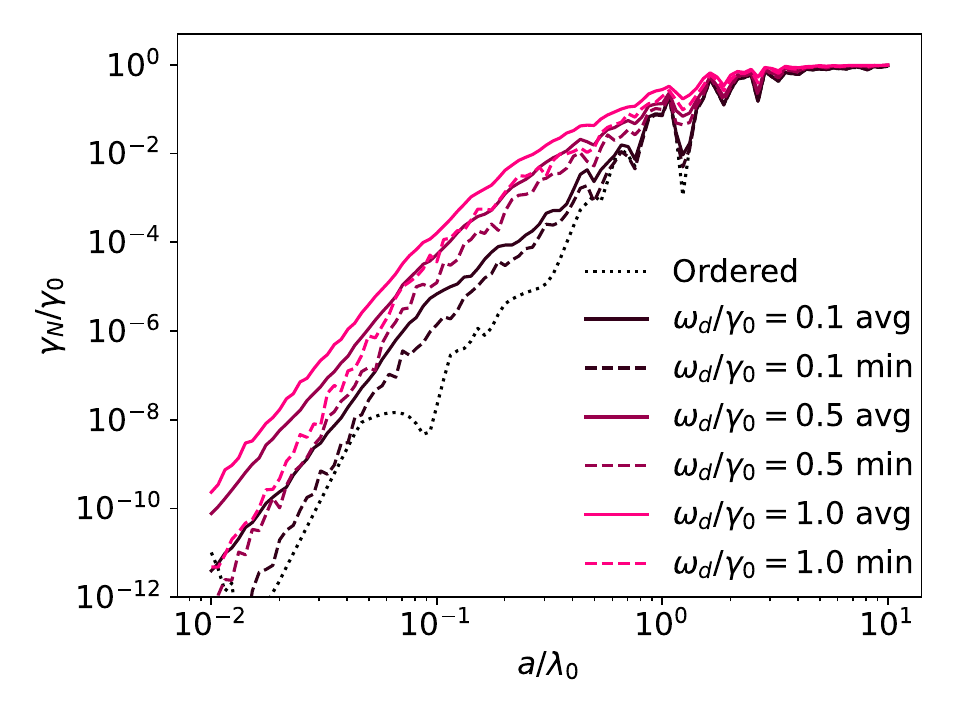}
		\label{fig:3d free detuning}
	\end{subfigure}

	\caption{The slowest decay rate $\gamma_N$ in the spectrum for various strengths of detuning disorder in the $N=50$ half waveguide (a), $N=50$ 1D free space chain (b), $N = 10 \times 10$ 2D free space square (c), and $N = 5 \times 5 \times 5$ 3D free space cube (d). In the half waveguide, spectral localization halts transport and still localizes energy within the bulk, suppressing decay. In free space, since atoms can no longer group together in space, dark singlets are no longer created, and disorder offers no advantage on average at any lattice spacing or strength.}
	\label{fig:detuning disorder}
\end{figure*}

For completeness, we also consider disorder in the transition frequency of each emitter, effectively applying random detunings to each atom. This is the case considered in Ref.~\cite{Rubies_Bigorda_2023}, where a speedup in emission was observed from detuning disorder in $N=3$ 1D chains in free space, and Ref.~\cite{Chen_2022}, where disorder in the totally symmetric Dicke Hamiltonian was used to slow decay. In Eq.~\eqref{eq:H free}, we restrict $\mathbf{G}(\mathbf{r}, \omega)$ to resonance, where $\omega = \omega_0$. To ensure this approximation is still valid, we restrict the strength of detuning to a single linewidth, i.e. $\omega_d \leq \gamma_0$. We define $\omega_d$ such that the detuning of each atom $\Delta_i = \omega_i - \omega_0$ is drawn from a uniform distribution of width $\omega_d$, so $\Delta_i \in [-\omega_d/2,\omega_d/2]$.

Since these small detunings will begin to break the resonant hopping between atoms, we still expect at least some spectral localization. However, because we can no longer push two atoms close together, we won't see any dark singlets or bright symmetric triplets. We therefore shouldn't expect to see any average suppression of decay in free space at any lattice spacing or disorder strength. In the half waveguide, however, where transport plays a dominant role in decay, we should still expect to see longer-lived states when disorder is present. In Fig.~\ref{fig:detuning disorder}, we plot the analog of Fig.~\ref{fig:hwg lattice spacing}a for detuning disorder in a half waveguide and 1D, 2D, and 3D arrays in free space, using the same parameters from the body of the paper. 

In the half waveguide, detuning disorder still localizes eigenstates and halts resonant hopping. We therefore observe a slow-down in decay for different detuning strengths as we'd expect. While spatial disorder showed equal suppression of decay for different disorder strengths at large $a$, here the opposite seems to be happening. In a very dense array, all different detuning disorder strengths suppress decay equally, whereas dilute arrays show a stronger dependence on $\omega_d$. The exact reason for this is left to future work, but we still observe a link between spectral and dynamical localization in the half waveguide. In free space, we see just what we expect: since dark singlets and similar few-body dark states cannot be created now, disorder doesn't suppress average decay at any lattice spacing or disorder strength.

\end{document}